\documentclass[12pt]{article}
\usepackage{graphicx}
\begin{document}
\rightline{NKU-09-SF2}
\bigskip
\begin{center}
{\Large\bf Decay of massless Dirac field around the  Born-Infeld Black Hole}

\end{center}
\hspace{0.4cm}
\begin{center}
Sharmanthie Fernando \footnote{fernando@nku.edu} \\
{\small\it Department of Physics \& Geology}\\
{\small\it Northern Kentucky University}\\
{\small\it Highland Heights}\\
{\small\it Kentucky 41099}\\
{\small\it U.S.A.}\\

\end{center}

\begin{center}
{\bf Abstract}
\end{center}

\hspace{0.7cm} 
In this paper we investigate the perturbations  by a massless Dirac spinor of a Born-Infeld black hole. The decay rates of the spinor field which is given by the quasinormal mode frequencies are computed using the sixth order WKB approximation. The behavior of the quasi normal modes with the non-linear parameter, temperature and charge of the black hole are analyzed in detail. We also compare the decay rates of the Born-Infeld black  hole  with its linear counter-part Reissner-Nordstrom black hole.


\section{Introduction}

When black holes are perturbed, the resulting oscillations lead to quasinormal modes (QNM). Such modes are defined as the solutions to the related wave equations characterized by purely ingoing waves at the horizon. In addition, one has to impose boundary conditions on the solutions at the asymptotic regions as well. In asymptotically flat space-times, the second boundary condition is for the solution to be purely outgoing  at spatial infinity. Once these boundary conditions are imposed, the resulting frequencies become complex and discrete. The frequencies of QNM's depend on the parameters of the black holes such as mass, charge and angular momentum and are independent of initial perturbations. If the radiation due to QNM modes are  detected in the future by gravitational wave detectors, it would be a clear way of identifying the possible charges of black holes. There are extensive studies of QNM's in various black-hole backgrounds in the literature. See the review by Kokkotas et. al. \cite{kok1} for more information. 

An important aspect of  QNM studies have been related to the conjecture relating anti-de Sitter(AdS) and conformal field theory (CFT) \cite{aha}. It is conjectured that  the imaginary part of the QNM's, which gives the time scale to decay the black hole perturbations, corresponds to the time scale of the CFT on the boundary to reach thermal equilibrium. There are many work on AdS black holes in four and higher dimensions on this subject and we will refer a few among them in \cite{horo} \cite{car1}  \cite{wang}  \cite{kon1} .

Another reason for intensive research on computing the QNM's has been 
due to Loop quantum gravity. The claims relating the real part of the high overtones of QNM's to entropy in loop quantum gravity has led to many works
such as  \cite{hod} \cite{corichi} \cite{mot1} \cite{dreyer} \cite{jo1} \cite{jo2} \cite{mot2} \cite{van}  \cite{set1} \cite{set2} \cite{kun} \cite{ander}.

QNM's of gravitational and scalar perturbations are extensively studied. In contrast, the number of works on spinor perturbations around black holes are smaller. QNM's of spinors of the Schwarzschild black hole (with and without the cosmological constant)  were studied by numerous authors. Some of them are the works by Zhidenko \cite{zhe}, Cho \cite{cho}, Shu et.al. \cite{shu1}, Jing \cite{jing1} \cite{jing2}. The Reissner-Nordstrom black hole ( with and without the cosmological constant) were studied by Zhou \cite{zhou}, Jing et.al. \cite{jing3} \cite{jing4} and Zhang et.al. \cite{zhang}. The Dirac field around the charged string black hole was studied by Shu et.al. \cite{shu2}. Neutrino QNM's of a Kerr-Newmann-de-Sitter black hole were calculated by Chang and Shen in \cite{chang}.

In this work we study  the massless spinors around the Born-Infeld black hole. Born-Infeld electrodynamics was first introduced in 1930's to obtain a classical theory of charged particles with finite self-energy \cite{born}. In Maxwell's theory of electrodynamics, the field of a point-like charge is singular at the origin and its energy is infinite. In contrast, in the Born-Infeld electrodynamics, the electric field of a point like object which is given by $E_r = Q/ \sqrt{r^4 + \frac{Q^2}{\beta^2}}$ is regular at the origin and its total energy is finite. Born-Infeld theory has received renewed attention since it turns out to play an important role in string theory. It arises naturally in open superstrings and in D-branes \cite{leigh}. The low energy effective action for an open superstring in loop calculations lead to Born-Infeld type actions \cite{frad}. It has also been observed that the Born-Infeld action arises as an effective action governing the dynamics of vector-fields on D-branes \cite{tsey1}. For a  review of aspects of Born-Infeld theory in string theory see Gibbons \cite{gib1} and Tseytlin \cite{tsey2}.

If  we construct black hole solutions to Einstein-Born-Infeld gravity, it can be observed that the non-singular nature of the electric field at the origin changes the structure of the space-time drastically. The singularity at the origin is dominated by the mass term $M/r$ rather than the charge term $Q^2/r^2$ as in the black holes of Einstein-Maxwell gravity \cite{chandra}. Here the possibility exists for $M$ to be positive, negative or zero leading to space-like or time-like singularities in contrast to the Reisser-Nordstrom black hole. Overall, the
physical properties of black holes change drastically due to the non-linear nature of electrodyanmics and is worthy of study. In this paper, we are focusing on the static charged black holes of the Born-Infeld electrodynamics. This is the non-linear generalization of the Reissner-Nordstrom black hole and is characterized by charge $Q$, mass $M$ and  the non-linear parameter $\beta$. The Born-Infeld black hole was obtained by Garcia et.al. \cite{Garcia} in 1984. Two years later, Demianski \cite{demia} also presented a solution known as EBIon which  differs with the one in \cite{Garcia} by a constant. Trajectories of test particles in the static charged Born-Infeld black hole were discussed by Breton \cite{nora1}. This black hole in isolated horizon framework was discussed by the same author in \cite{nora2}. Gibbons and Herdeiro \cite{gib2} derived a Melvin Universe type solution describing a  magnetic field permeating the whole Universe in Born-Infeld electrodynamics coupled to gravity. By the use of electric-magnetic duality, they also obtained Melvin electric and dyonic Universes.

The paper is presented as follows: In section 2, the Born-Infeld black hole solutions are introduced. In section 3, the spinor perturbations are given. In section 4, we will computer the QNM's and discuss the results. Finally, the conclusion is given in section 5.

\section{Introduction to the Born-Infeld black hole}

In this section, we introduce the Born-Infeld black hole. The action for the Einstein gravity coupled to Born-Infeld electrodynamics is given by,
\begin{equation}
S = \int d^4x \sqrt{-g} \left[ \frac{R }{16 \pi G} + L(F) \right]
\end{equation}
Here, $L(F)$ is a function of the field strength $F_{\mu \nu}$ and  for Born-Infeld electrodynamics may be expanded to be,
\begin{equation}
L(F) = 4 \beta^2 \left( 1 - \sqrt{ 1 + \frac{ F^{\mu \nu}F_{\mu \nu}}{ 2 \beta^2}} \right)
\end{equation}
Here, $\beta$ has dimensions $length^{-2}$ and $G$  $length^2$. In the following sections it is assumed that $16 \pi G = 1$. Note that when $\beta \rightarrow \infty$, the Lagrangian $L(F)$ approaches the one for Maxwell's  electrodynamics given by $ - F^2$.

The static charged black hole with spherical symmetry can be obtained as,
\begin{equation}
ds^2 = -f(r) dt^2 + f(r)^{-1} dr^2 + r^2 ( d \theta^2 + Sin^2(\theta) d \varphi^2)
\end{equation}
with,
\begin{equation}
f(r) = 1 - \frac{2M}{r} + \frac{2 \beta^2 r^2}{3} \left( 1 - \sqrt{ 1 + \frac{Q^2}{r^4 \beta^2}} \right) + \frac{ 4 Q^2}{ 3 r^2} \hspace{0.2cm}   _2F_1 \left( \frac{1}{4}, \frac{1}{2}, \frac{5}{4}, -\frac{Q^2}{ \beta^2 r^4} \right)
\end{equation}
Here $_2F_1$ is the hypergeometric function.
In the limit $\beta \rightarrow \infty$, the elliptic integral can be expanded to give,
\begin{equation}
f(r)_{RN} = 1 - \frac{2 M}{r} + \frac{ Q^2}{r^2}
\end{equation}
resulting  in the function $f(r)$ for the Reissner-Nordstrom black hole for Maxwell's electrodynamics.
Near the origin, the function $f(r)$ has the behavior,
\begin{equation}
f(r) \approx 1 - \frac{( 2M - A)}{r} - 2 \beta Q + \frac{2 \beta^2}{3} r^2  + \frac{ \beta^3}{5}  r^4
\end{equation}
Here,
\begin{equation}
A = \frac{1}{3}\sqrt{ \frac{\beta}{ \pi} } Q^{3/2} \Gamma \left(\frac{1}{4} \right)^2
\end{equation}
Depending on the values of $M$, $Q$ and $\beta$, the function $f(r)$ for the Born-Infeld black hole can  have two roots, one root or none. When $f(r)$ has two roots, the behavior
of $f(r)$  is similar to the Reissner-Nordstrom black hole. When it has only a single root, the black hole behave similar to the Schwarzschild black hole. Hence the Born-Infeld black  hole is  interesting  since it possess  the characteristics of the most well known black holes in the literature. 

Extreme black holes are possible when both  $f(r) =0$ and $f'(r) =0$. This condition yields the horizon radii as,
\begin{equation}
r_{ex} = \frac{\sqrt{ 4 \beta^2 Q^2 - 1}} { 2 \beta}
\end{equation}
Hence, extreme black holes are possible only if $ Q \beta > 1/2$. We will plot the graphs for $f(r)$ in the following figure. Note that for $ Q \beta < 1/2$ only Schwarzschild type black holes are possible.

\newpage

\begin{center}
\scalebox{.9}{\includegraphics{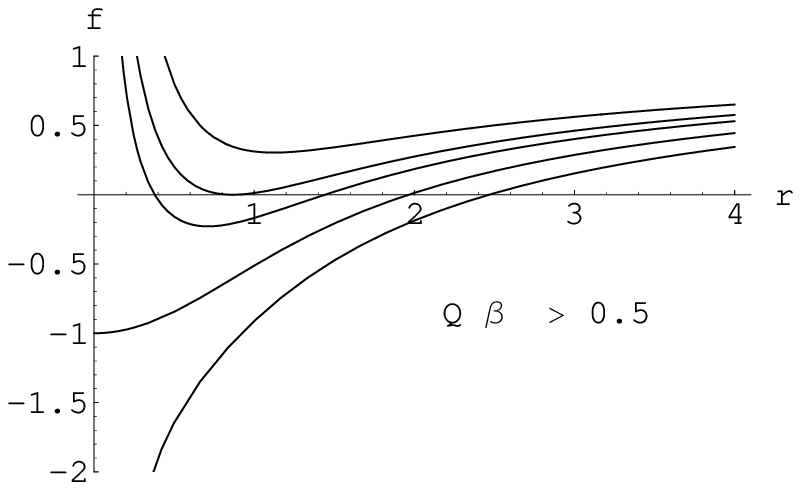}}\\

\scalebox{.9}{\includegraphics{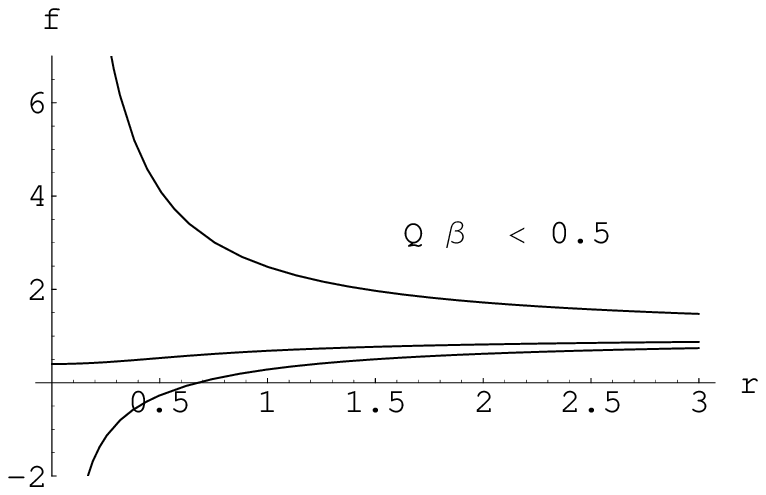}}

\vspace{0.3cm}
\end{center}
Figure 1. The behavior of the function $f(r)$ when the mass is changed is plotted for fixed values of $Q$ and $ \beta$. In the top graph $Q =1$ and $\beta = 1$. In the  bottom graph $ Q =0.3$ and $\beta =1$.

A detailed description of the characteristics of the Born-Infeld black hole is given \cite{rasheed2} and \cite{nora1}. The Hawking temperature of the black hole is given by,
\begin{equation}
T = \frac{1}{4\pi} \left[ \frac{1}{r_+}  + 2\beta \left( r_+ \beta - \frac{\sqrt{(Q^2 + r_+^2 \beta^2)}}{r_+} \right) \right]
\end{equation}
Here, $r_+$ is the event horizon of the black hole which is a solution of $f(r)=0$. The zeroth and the first law of black holes in Born-Infeld electrodynamics are discussed in detail in \cite{rasheed2}. Static charged black hole solution to the above action with a cosmological constant was presented in \cite{fer1} \cite{cai} and was extended to higher dimensions in \cite{dey}.

\section{Massless spin $\frac{1}{2}$ field  around spherically symmetric static black hole}

In this section, we will develop the equations for a  spin 1/2 field  in the background of a  static spherically symmetric space-time. The general equation for a massless Dirac spinor field in curved space-time can be written as,
\begin{equation}
\gamma^a e^{\mu}_{a} ( \partial_{\mu} - \Gamma_{\mu} ) \psi =0
\end{equation}
Here, $e^{\mu}_{a}$ are the inverse of the tetrads $e^{\mu}_{a}$. The metric tensor $g_{\mu \nu}$ of the space-time considered and the tetrad $e^{a}_{\mu}$ are related by,
\begin{equation}
g_{\mu \nu} = \eta_{a b} e^{a}_{\mu} e^{b}_{\nu}
\end{equation}
Here, $\eta_{a b} = ( -1,1,1,1)$  is  the metric in the flat space. The $\gamma^a$ matrices are defined by,
\begin{eqnarray}
\gamma^{0} =  \left(\begin{array}{cc} i & 0 \\ 0 & - i
\end{array} \right) ; \;\;\;\; 
\gamma^{a} =  \left(\begin{array}{cc}  0  & i  \sigma^a \\  - i  \sigma^a & 0
\end{array} \right)
\end{eqnarray}
Here, $ \sigma^{a}$ are the Pauli spinors 
\begin{eqnarray}
\sigma^{1} =  \left(\begin{array}{cc} 1 & 0 \\ 0 & -1
\end{array} \right) ; \;\;\;\; 
\sigma^{2} =  \left(\begin{array}{cc}  0 & - i \\ i & 0
\end{array} \right);  \;\;\;\; 
\sigma^{3} = \left(\begin{array}{cc} 0 & 1 \\ 1 & 0
\end{array} \right)
\end{eqnarray}
The $\gamma^a$ matrices satisfy the anti-commuting relations,
\begin{equation}
\{ \gamma^a, \gamma^b \} = 2 \eta^{a b} I
\end{equation}
The spin connection $\Gamma_{\mu}$ is defined as,
\begin{equation}
\Gamma_{\mu} = - \frac{1}{8} [ \gamma^a , \gamma^b ] \Lambda_{a b \mu}
\end{equation}
where,
\begin{equation}
\Lambda_{a b \mu} = e^{\nu}_{a} e_{b \nu; \mu}
\end{equation}
$e_{ b \nu ; \mu}$ is the covariant derivative of $e_{ b \nu}$ given by,
\begin{equation}
e_{ b \nu ; \mu} = \partial_{\mu} e_{b \nu} - \Gamma_{\mu \nu}^{\alpha} e_{b \alpha}
\end{equation}
$\Gamma_{ \nu \mu} ^{\alpha} $ are the Christoffel symbols.

We will compute the Dirac equation  for a space-time which is spherically symmetric and static given by,
\begin{equation}
ds^2 = - A(r)^2 dt^2 +  B(r)^2 dr^2 + R(r)^2 ( d \theta^2 + sin^2(\theta) d \varphi^2)
\end{equation}
The tetrads for this space-time are given by,
\begin{equation}
e^a_{\mu} = diag \left( A, B, R, R sin \theta \right)
\end{equation}
The non-zero components of $\Lambda_{a b \mu}$ defined in eq.(16) for the above metric are computed as,
$$ \Lambda_{ 0 1 t} = - \frac{A'}{B}$$
$$ \Lambda_{ 1 2 \theta} = - \frac{ R'} { B}$$
$$ \Lambda_{1 3 \varphi } = - sin \theta \frac{ R'} { B}$$
\begin{equation}
\Lambda_{2 3 \varphi } = - cos \theta
\end{equation} 
Therefore, the related spin connections becomes,
$$ \Gamma_t = \frac{ A'}{ 2 B} \gamma^ 0 \gamma^ 1$$
$$ \Gamma_r = 0$$
$$ \Gamma_{\theta} = \frac{ R'}{2 B} \gamma^1 \gamma^2$$
\begin{equation}
\Gamma_{\varphi} = \frac{cos \theta }{2} \gamma^2 \gamma^3 + \frac{ sin \theta R'}{2 B} \gamma^1 \gamma^3
\end{equation}
The Born-Infeld black hole considered in this paper has the functions
$A = B^{-1} $ and $R = r$. For further simplification, we will take $ A = B^{-1} = \sqrt{f(r)}$. The function $\psi$ is redefined as,
\begin{equation}
\psi = \frac{\Phi}{ f^{1/4} }
\end{equation}
Then the Dirac equation simplifies to,
\begin{equation}
-\frac{\gamma_0}{\sqrt{f}} \frac{\partial \Phi}{\partial t} + \sqrt{f} \gamma_1 
\left( \frac{\partial}{\partial r} + \frac{1}{r} \right) \Phi +
\frac{\gamma_2}{r} \left( \frac{\partial}{\partial \theta} + \frac{ cot \theta} {2} \right) \Phi + \frac{ \gamma_3} { r sin \theta} 
\frac{ \partial \Phi} { \partial \varphi} = 0
\end{equation}
Before looking for solutions for the above equation, we will make another assumption. The field considered here is massless and has spin $\frac{1}{2}$. Hence the allowed solutions to the Dirac equation are circularly polarized. This is discussed in \cite{lee} and \cite{brill} at length. According to \cite{brill}, the above spinors have  right handed circular polarization. Hence, the allowable spin states satisfy the condition,
\begin{equation}
( 1 - i \gamma_5 ) \Phi = 0
\end{equation}
where
\begin{equation}
\gamma^5 = \gamma^0 \gamma^1 \gamma^2 \gamma^3
\end{equation}
Then, eq.(24) will be satisfied by,
\begin{eqnarray}
\Phi =  \left(\begin{array}{l} \Phi_1( t,r,\theta, \varphi) \\ 
\Phi_2( t,r,\theta, \varphi) \\ 
\Phi_1 ( t,r,\theta, \varphi) \\ 
\Phi_2 ( t,r,\theta, \varphi)
\end{array} \right) 
\end{eqnarray}
Now, the Dirac equation yield two identical set of equations each coupling $\Phi_1$ and $\Phi_2$. Hence only one set of the equations will be solved. Let us redefine the  components  $\Phi_1$ and $\Phi_2$ as follows:
\begin{eqnarray}
\left(\begin{array}{l} \Phi_1( t,r,\theta, \varphi) \\ 
\Phi_2( t,r,\theta, \varphi) 
\end{array} \right) = \left(\begin{array}{l}   \frac{i F(r)}{r} H_1(\theta, \varphi) \\ 
 \frac{G(r)}{r} H_2(\theta, \varphi) 
\end{array} \right) e^{- i \omega t}
\end{eqnarray}
Substituting the above in eq.(23) yields,
\begin{equation}
\left( \frac{ i \omega r}{ \sqrt{f} } F -  r \sqrt{f} \frac{ dF }{ dr}\right) \frac{1}{G} + \left( \frac{cot \theta}{2} H_2 + \frac{i}{sin \theta} \frac{ \partial H_2} {\partial \varphi} + \frac{\partial H_2} {\partial \theta} \right) \frac{1}{ H_1}=0
\end{equation}

\begin{equation}
\left( \frac{  i \omega r}{ \sqrt{f} } G +  r \sqrt{f} \frac{ dG}{ dr} \right) \frac{1}{F} + \left( \frac{cot \theta}{2} H_1 - \frac{i}{sin \theta} \frac{ \partial H_1} {\partial \varphi} + \frac{\partial H_1} {\partial \theta} \right) \frac{1}{ H_2}=0
\end{equation}
To solve the angular part of the equation the following operators are defined,
\begin{equation}
\partial_{+} = - \left( \frac{cot \theta}{2}  + \frac{i}{sin \theta} \frac{ \partial } {\partial \varphi} + \frac{\partial} {\partial \theta} \right) 
\end{equation}

\begin{equation}
\partial_{-} = - \left( \frac{cot \theta}{2}  + \frac{i}{sin \theta} \frac{ \partial } {\partial \varphi} + \frac{\partial} {\partial \theta} \right) 
\end{equation}
These operators were discussed in \cite{gold} at length. It was shown that the operators act on spin weighted spherical harmonics $_sY_{lm}$ as ladder operators. In particular if the spin $ s = \frac{1}{2}$, then the above operators gives the following relations;
\begin{equation}
\partial_+ ( _{-\frac{1}{2}}Y_{lm} ) =  ( l + \frac{1}{2} )  _{\frac{1}{2}}Y_{lm} 
\end{equation}
\begin{equation}
\partial_- ( _{\frac{1}{2}}Y_{lm} ) = - ( l + \frac{1}{2} )  _{-\frac{1}{2}}Y_{lm} 
\end{equation}
By comparing the angular parts of the equation of the Dirac equation, one can see that  $H_1$ and $H_2$ are the spin $\frac{1}{2}$ weighted spherical harmonics given by

\begin{equation}
H_1 = _{\frac{1}{2}}Y_{lm}
\end{equation}
\begin{equation}
H_2 = _{-\frac{1}{2}}Y_{lm}
\end{equation}
Hence the angular part of the eq.(28) and eq.(29) becomes $\lambda = l + \frac{1}{2}$. Note that $l$  can be written as $l = (2 k - 1)/2$ where $k$ is a positive integer. Also $ -l \leq m \leq l$. Therefore, the radial part simplifies to,
$$d_{*} G - i \omega G + W(r)  F = 0$$
\begin{equation}
d_{*} F + i \omega F + W(r)  G = 0
\end{equation}
$r_*$ is the well known ``tortoise'' coordinate given by,
\begin{equation}
dr_{*} = \frac{dr}{f}
\end{equation}
The function $W(r)$ is,
\begin{equation}
W(r) = \frac{\lambda \sqrt{f}}{r}
\end{equation}
Two new functions $Z_{\pm}$ are defined as,
\begin{equation}
Z_{\pm} = F \pm G
\end{equation}
and  eq.(36) can be decoupled as,
\begin{equation}
d_{*}^2 Z_{\pm} + ( \omega^2 - V_{\pm} ) Z_{\pm} = 0
\end{equation}
Here, $V_{\pm}$ are related to $W(r)$ as,
\begin{equation}
V_{\pm} = \pm d_{*} W  + W^2
\end{equation}
The effective potential $V_+$ for the Born-Infeld black hole is plotted to show how it changes with charge $Q$ and the non-linear parameter $\beta$ in the following figures.
\begin{center}
\scalebox{.9}{\includegraphics{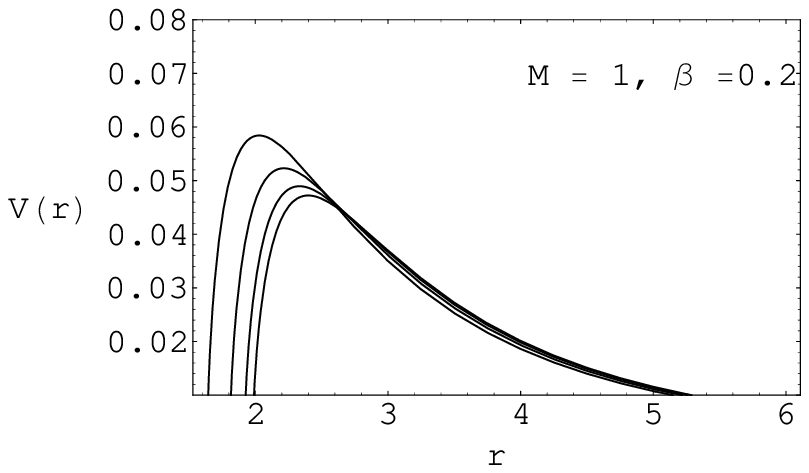}}

\vspace{0.3cm}
\end{center}
Figure 2. The behavior of the effective potential $V_+$ with 
the charge for the Born-Infeld black hole. Here, $M=1$, $\beta=0.2$ and $\lambda=1$. The height of the potential decreases when the charge decreases.

\begin{center}
\scalebox{.9}{\includegraphics{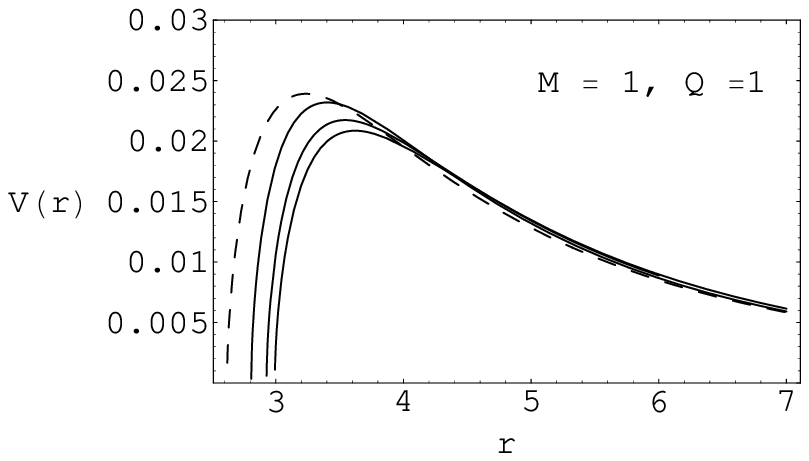}}

\vspace{0.3cm}
\end{center}

Figure 3. The behavior of the effective potential $V_+$ with 
the non-linear parameter $\beta$. Here, $M=1$, $Q=1$ and $\lambda=1$. The maximum height of the potential increases as $\beta$ increases. The dashed one is the potential for the Reissner-Nordstrom black hole with same mass and charge.

\subsection{Comparison with other fields}

Gravitational  perturbation and the scalar perturbations 
of the Born-Infeld black hole were studied by Fernando in \cite{fer2}\cite{fer3}.
The equations were shown to be simplified to a Schrodinger type equation as for the spinor 1/2 field studied here.  Here we will plot the effective potentials for the three fields to compare the behavior.
\begin{center}
\scalebox{.9}{\includegraphics{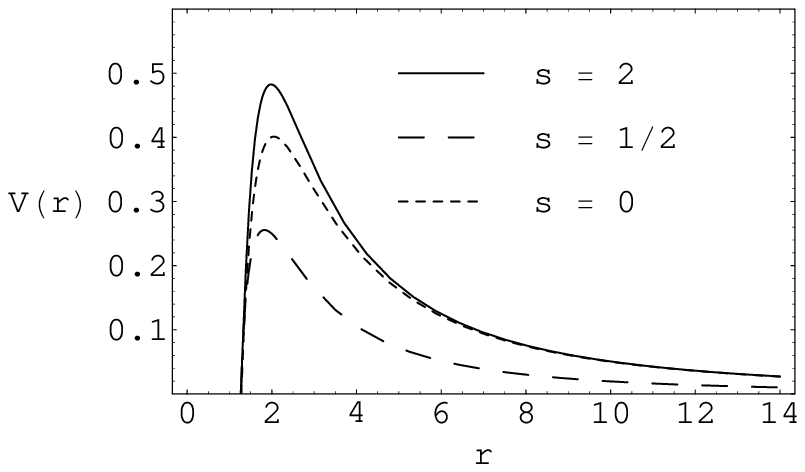}}

\vspace{0.3cm}
\end{center}

Figure 4. The behavior of the effective potential $V_+$ for fields of spin 2, 1/2 and 0.  Here, $M=1$, $Q=1$, $\beta=0.4$. The value of $l=2$ for scalar and gravitational potential, while $\lambda = 2$ for the spinor potential.

\section{Quasi-normal modes of the Born-Infeld black hole}

Usually, the fundamental equation of black hole perturbations given in eq.(40) cannot be solved analytically. This is the case for almost all black holes in 3+1 dimensions. In 2+1 dimensions, there are two black hole solutions ( BTZ black hole \cite{bir1}  and the charged dilaton black hole \cite{fer4}) which can be solved to give exact values of QNM's. In five dimensions, exact values are obtained for vector perturbations by Nunez and Starinets \cite{nun}.

There are several approaches  to compute QNM's in the literature. Here, a semi analytical technique developed by Iyer and Will \cite{will} is followed. The method makes use of the WKB approximation. The formula developed by Iyer and Will was for up to third order. Following an approach suggested by Zaslavskii \cite{zal}, Konoplya presented the sixth order WKB formula in \cite{kono2}. We will follow the formalism presented in that paper. The WKB formula is given by,
\begin{equation}
\frac{ \omega^2 - V_0}{ \sqrt{ - 2 V_0^{''}}} - L_2 - L_3 - L_4 - L_5 - L_6 = n + \frac{1}{2}
\end{equation}
Here, $V_0$ and $V_0^{''}$ represents the maximum potential and the second derivative with respect to the tortoise coordinates where the maximum occurs. $L_2$ and $L_3$ are given in \cite{will} and $L_3$, $L_4$, $L_5$ and $L_6$ are given in \cite{kono2}. $n$ is the overtone number. The $\omega$ is represented as  $\omega = \omega_R - i \omega_I$ and presented accordingly in the tables and the graphs. Unless states specifically, the QNM's are computed for $n =0$.

First, the quasi normal modes for $V_+$ are computed to see the behavior with the non-linear parameter  $\beta$ and graphed in the following figures.

\begin{center}
\begin{tabular}{|l|l|l|l|l|r} \hline \hline
 $\beta$ & $\omega_R$  &  $\omega_I$ \\ \hline
0.001 & 0.291904 & 0.0489188 \\ \hline
0.002 & 0.293482 & 0.0491336  \\ \hline
0.003 & 0.294531 & 0.0492596  \\ \hline
0.004 & 0.295312 & 0.0493408  \\ \hline
0.005 & 0.295927 & 0.0493945  \\ \hline
0.008 & 0.297194 & 0.0494625 \\ \hline
0.01 & 0.297757 & 0.0494647  \\ \hline
0.015 & 0.298665 & 0.0494079  \\ \hline
0.02 & 0.299191 & 0.0493255  \\ \hline

\end{tabular}
\end{center}

\vspace{0.3cm}

\vspace{0.3cm}
\begin{center}
\scalebox{.9}{\includegraphics{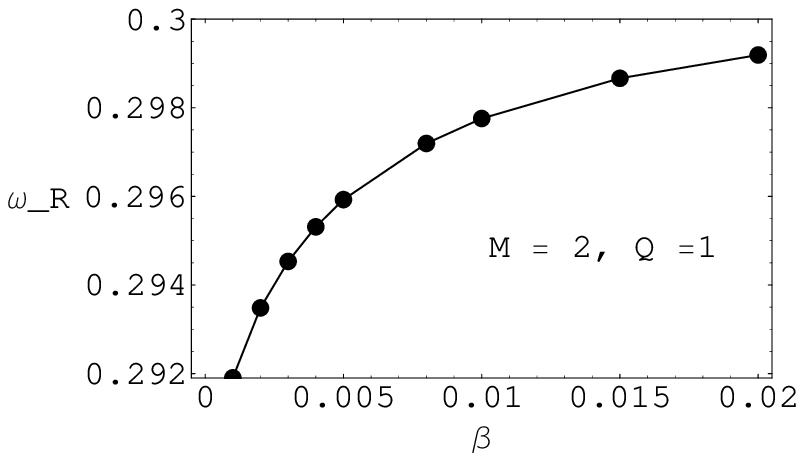}}
\vspace{0.3cm}
\end{center}

Figure 5. The behavior of Re $\omega$ with the non-linear parameter $\beta$ for $M=2$, $Q=1$ and $\lambda=3$.

\newpage
\begin{center}
\scalebox{.9}{\includegraphics{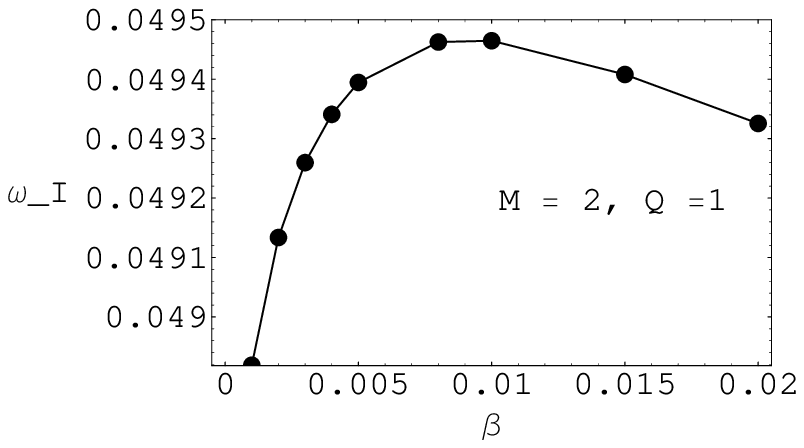}}
\vspace{0.3cm}
\end{center}

Figure 6. The behavior of Im $\omega$ with the non-linear parameter $\beta$ for $M=2$, $Q=1$ and $\lambda=3$
\\

The spinor perturbation decay faster when the non-linear parameter $\beta$ increases. However, beyond a certain value, the decay becomes slower as is evident from the graph.

Next, we have computed the QNM's by varying the charge of the black hole. We also computed the QNM's for the Reissner-Nordstrom black hole with the same mass and the charge. The results are given in the following table and the figures.

\vspace{0.4cm}

\begin{center}
\begin{tabular}{|l|l|l|l|r|} \hline \hline
 $Q$ & $\omega_R$(BI)  &  $\omega_I$(BI) & $\omega_r$ (RN) & $\omega_I$ (RN) \\ \hline

0.1 & 0.57506 & 0.0963636 & 0.575063 & 0.0963601\\ \hline
0.2 & 0.57796 & 0.0965668 & 0.578015 & 0.0965175 \\ \hline
0.3 & 0.582785 & 0.0969885 & 0.583084 & 0.0967732\\ \hline
0.4 & 0.589571 & 0.0976722 & 0.59052 & 0.0971135 \\ \hline
0.5 & 0.598382 & 0.0986511 & 0.600731 & 0.0975103 \\ \hline
0.6 & 0.609334 & 0.0999679 & 0.614369 & 0.0979046 \\ \hline
0.7 & 0.622627 & 0.10166 & 0.632506 & 0.0981638 \\ \hline
0.8 & 0.638537 & 0.103767 & 0.657033 & 0.0979584 \\ \hline
0.9 & 0.679767 & 0.109443 & 0.746085 & 0.0883246 \\ \hline

\end{tabular}
\end{center}

\begin{center}
\vspace{0.3cm}

\scalebox{.9}{\includegraphics{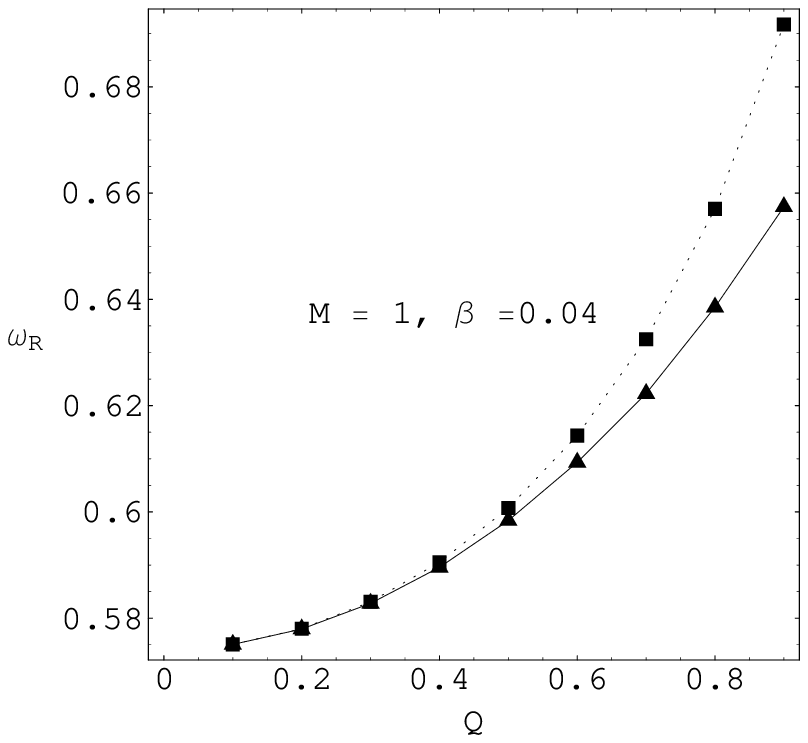}}

\vspace{0.2cm}
\end{center}

Figure 7. The behavior of Re $\omega$ with the charge $Q$ for $M=1$, $\beta=0.04$ and $\lambda =3$. The dotted lines show the graph for Reissner-Nordstrom and the dark lines show the one for the Born-Infeld black hole.

\begin{center}
\scalebox{.9}{\includegraphics{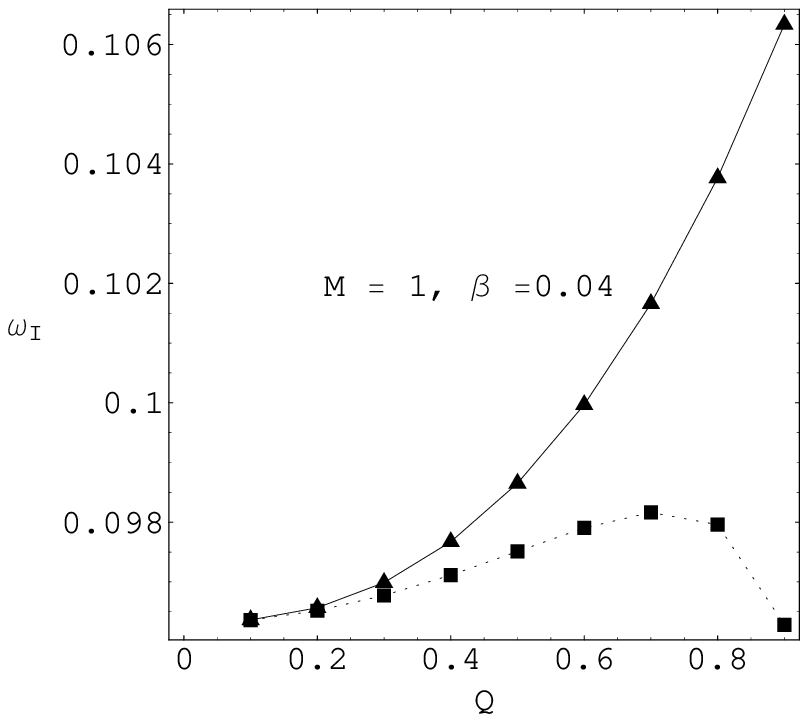}}

\vspace{0.3cm}

\end{center}

Figure 8. The behavior of Im $\omega$ with the charge $Q$ for $M=1$, $\beta=0.04$ and $\lambda =3$. The dotted lines show the graph for Reissner-Nordstrom and the dark lines show the one for the Born-Infeld black hole.
\\

One can observe that the decay of the spinor field in Born-Infeld black hole is faster than the Reissner-Nordstrom black hole for this particular values of the parameters. For scalar field perturbations the decay for the Born-Infeld black hole is also  faster than the Riessner-Nordstrom black hole \cite{fer3}. Note that, for the Reissner-Nordstrom black holes, $Q=1$, $M=1$ represents the extreme black hole. This will explain why the $\omega_I$ is approaching zero around $Q=0.9$ in the table.

The behavior of the quasi normal modes for the Born-Infeld black hole with varying $\lambda$  was also studied as given in the following table.
\begin{center}
\begin{tabular}{|l|l|l|r} \hline \hline
 $\lambda$ & $\omega_R$  &  $\omega_I$ \\ \hline
2 & 0.198638 & 0.0491501 \\ \hline
3 & 0.299979 & 0.049059 \\ \hline
4 & 0.400892 & 0.0490351 \\ \hline
5 & 0.501645 & 0.0490251 \\ \hline
6 & 0.60232 & 0.0490197 \\ \hline
7 & 0.702949 & 0.0490165 \\ \hline
8 & 0.803551 & 0.0490144 \\ \hline

\end{tabular}

\vspace{0.3cm}

\scalebox{.9}{\includegraphics{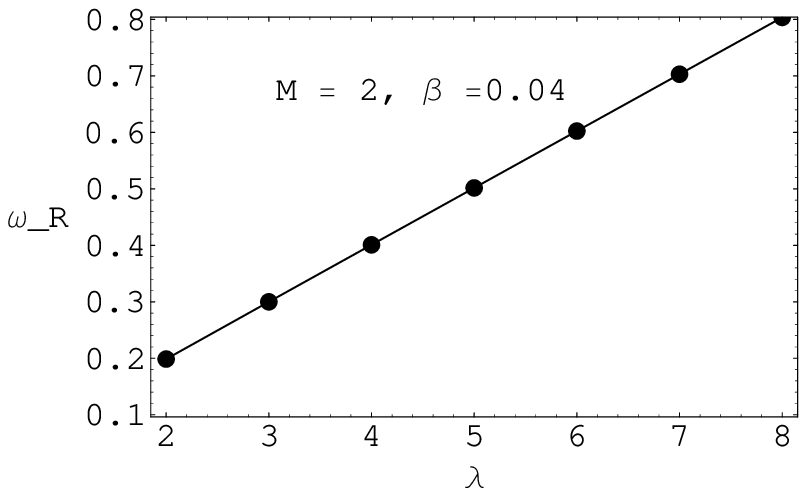}}

\vspace{0.3cm}

\end{center}
Figure 9. The behavior of Re $\omega$ with the $\lambda$ for $M=2$, $\beta=0.04$ and $Q=1$

\begin{center}
\scalebox{.9}{\includegraphics{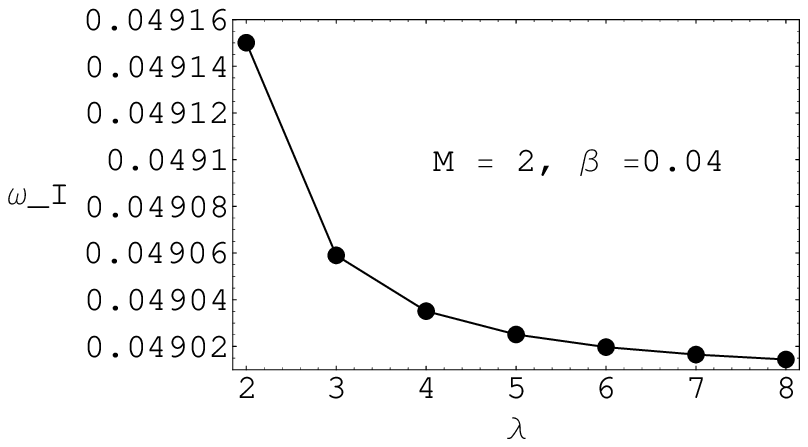}}

\vspace{0.3cm}

\end{center}
Figure 10. The behavior of Im $\omega$ with the $\lambda$  for $M=2$, $\beta=0.04$ and
$Q = 1$\\

The real part of $\omega$ shows a linear relation with $\lambda$ while the imaginary part decreases to a stable limit.

Next, the behavior of the Im $\omega$ with temperature of the black hole is  studied. The plot of $Im (\omega)$ vs temperature is given in the Figure 11.
One can see a linear behavior of Im $\omega$ with the temperature. This behavior is similar to the Schwarzschild-anti-de-Sitter black hole studied by Horowitz and Hubeny \cite{horo}.

\begin{center}
\vspace{0.3 cm}

\scalebox{.9}{\includegraphics{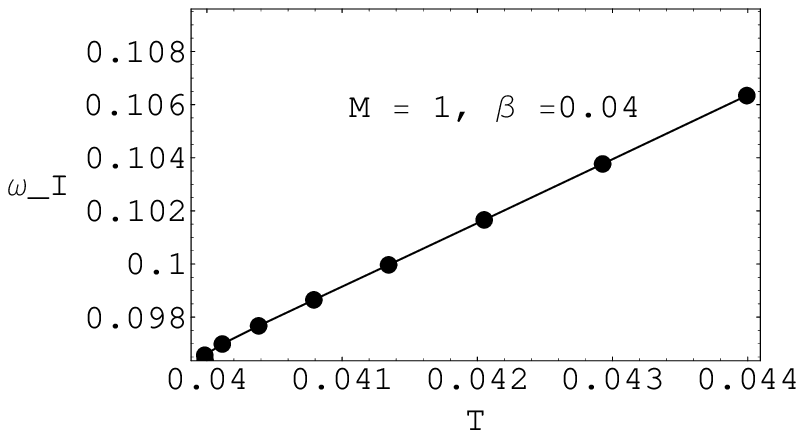}}

\vspace{0.3cm}
\end{center}

 Figure 11. The behavior of $\omega_I$ with the Temperature for  $\lambda=3$. The mass $M =1$ and $\beta = 0.04$.

The overtones of the QNM frequencies can be obtained only for $\lambda > n$ as stated in \cite{kono3}. The sixth order WKB method is used for $\lambda =8$, $ M =2$, $\beta =0.04$  and $Q =1$ for the first four overtones as follows:

\begin{center}
\begin{tabular}{|l|l|l|r} \hline \hline
 n & $\omega_R$  &  $\omega_I$ \\ \hline
1 & 0.800366 & 0.147305 \\ \hline
2 & 0.794092 & 0.246376 \\ \hline
3 & 0.784922 & 0.346724 \\ \hline
4 & 0.773144 & 0.448811 \\ \hline

\end{tabular}
\end{center}
\vspace{0.3cm}
The real part of the $\omega$ decreases for higher overtones, while the imaginary part increases with the overtone number $n$. This feature is similar to the Shwarzschild black hole as well as the charged black hole in Gauss-Bonnet black hole presented by Konoplya in \cite{kono2}.

\section{Conclusions}

We have studied the spinor perturbations of a Born-Infeld black  hole. The quasinormal mode frequencies  are computed using the sixth order WKB method. The parameters of the theory, charge $Q$, non-linear parameter $\beta$, $\lambda$ and overtone number $n$ are varied to compute the $\omega$. For all values considered, the value of the imaginary part of $\omega$ is negative leading to the decay of the spinor  field. In comparison with the Reissner-Nordstrom black hole, the decay rates are faster  for the Born-Infled black hole. It is interesting to note that for gravitational and scalar perturbations, the fields of the Born-Infeld black hole also decayed faster as noted in \cite{fer2} \cite{fer3}.

One can also conclude that the Born-Infeld black holes are stable for the domain of the parameters considered in this paper. However, since  we have done these computations for a Born-Infeld back hole which behave like the Schwarzschild black hole near the origin,  it is necessary to do an evaluation based on all the parameters to fully understand the  behavior of the $\omega_I$ and to make remarks on stability of the black hole. Perhaps one can find an analytical approach to find the stability properties of the black hole for all parameters considered. This may be an interesting possibility for future work.

A natural extension of this work is to study the stability of Born-Infeld-AdS black hole. The developments in string theory on  AdS/CFT duality and  the relation of Born-Infeld theory to string theory is a motivation for this.
The Reissner-Nordstrom-AdS black hole has been shown to be unstable for linear perturbations in \cite{gub}. 

It was noted in section 2 that the Born-Infeld black hole behaves in a similar fashion as the Reissner-Nordstrom for certain parameters of the theory. In particular, the possibility of extreme black holes exists. It is well known that the extreme Reissner-Nordstrom black hole is supersymmetric. It can be embedded in a N=2 supergravity theory \cite{gib3} \cite{gib4}. Quasinormal modes of these extreme black holes were studied by Onozawa et.al. \cite{ono} and was shown to have the same frequencies. It will be interesting to see any sign of such behavior for the Born-Infeld black hole.

\vspace{0.2cm}

{\bf Acknowledgments}: This work was supported in part by a grant from the Kentucky Space Grant Consortium. The author likes to thank R. Konoplya for providing the sixth order WKB {\it Mathematica} file. The author also  thank Don Krug for helping with {\it Mathematica} programming.

\end{document}